\documentclass[11pt,a4paper,oneside]{article}
\usepackage[utf8]{inputenc}
\usepackage[T2A,T1]{fontenc}
\usepackage{titlesec}
\usepackage{amsmath}
\usepackage{amsfonts}
\usepackage{mathtools}
\usepackage{amssymb}
\usepackage{graphicx}
\usepackage{subcaption}
\usepackage{amsthm}
\usepackage{setspace}
\usepackage{pgfplots}
\usepackage{fullpage}
\usepackage{exptex}
\usepackage{expthmi}
\usepackage{url}
\newcommand{\pkg}[1]{{\normalfont\fontseries{b}\selectfont #1}}
\let\proglang=\textsf
\let\code=\texttt
\usepackage[hidelinks]{hyperref}
\usepackage{rotating}
\usepackage{threeparttable}
\usepackage{multirow}
\usepackage{changepage}
\rmfamily
\makeindex
\makeatletter
\titleformat{\section}
{\normalfont\large\bfseries}{\thesection}{1em}{}

\providecommand{\keywords}[1]
{
  \small	
  \textbf{\textit{Keywords---}} #1
}

\usepackage[sorting=none]{biblatex}
\begin{document}

\title{Similarity network fusion for scholarly journals}
\author{Alberto Baccini\footnote{Department of Economics and Statistics, University of Siena}, Federica Baccini\footnote{Department of Informatics, University of Pisa}, Lucio Barabesi\footnote{Department of Economics and Statistics, University of Siena}, Yves Gingras\footnote{Université du Quebec, Montreal}, Mahdi Khelfaoui\footnote{Université du Quebec à Trois-Rivières}}
\author{Federica Baccini$^1$ \and Lucio Barabesi$^2$ \and Alberto Baccini$^2$\footnote{Alberto Baccini is the recipient of a grant by the Institute For New Economic Thinking Grant ID INO19-00023. The research is funded by the Italian Ministry of University, PRIN project: 2017MPXW98. The funders had no role in study design, data collection and analysis, decision to publish, or preparation of the manuscript. Comments of two referees are gratefully acknowledged.}\and Mahdi Khelfaoui$^3$\and Yves Gingras$^4$}
\date{%
    $^1$Department of Computer Science, University of Pisa, and IIT-CNR of Pisa, Italy\\%
    $^2$Department of Economics and Statistics, University of Siena, Italy\\%
    $^3$Université du Québec à Trois-Rivières, Canada\\%
    $^4$Université du Québec à Montreal, Canada\\[2ex]%
}
\maketitle

\begin{abstract}
\let\proglang=\textsf
\let\code=\texttt
This paper explores intellectual and social proximity among scholarly journals by using network fusion techniques. Similarities among journals are initially represented by means of a three-layer network based on co-citations, common authors and common editors. The information contained in the three layers is then combined by building a fused similarity network. The fusion consists in an unsupervised process that exploits the structural properties of the layers. Subsequently, partial distance correlations are adopted for measuring the contribution of each layer to the structure of the fused network. Finally, the community morphology of the fused network is explored by using modularity. In the three fields considered (i.e. economics, information and library sciences and statistics) the major contribution to the structure of the fused network arises from editors. This result suggests that the role of editors as gatekeepers of journals is the most relevant in defining the boundaries of scholarly communities. In information and library sciences and statistics, the clusters of journals reflect sub-field specializations. In economics, clusters of journals appear to be better interpreted in terms of alternative methodological approaches. 
\end{abstract}

\newpage
\section{Introduction}

The use of multilayer networks for representing and analyzing data is not common in scientometrics. It constitutes a major improvement with respect to the classical single layer approach, where only a single type of information about structural relations in a network is usually considered. In this paper a classical scientometric problem is tackled in a multilayer setting by using techniques imported from bio-informatics. 

The classical problem stems on the delineation of scientific fields and scholarly communities, by developing a fine grained classification of scholarly journals \cite{Borgman}. This problem is at the intersection of many different streams of scientometric literature, briefly recalled in Section 2. The focus of this paper is on a descriptive classification of journals aimed at representing intellectual and social structures of scientific fields through relationships between journals. Journals are classified by considering information contained in a three-layer network. The first layer contains information about journal co-citations organized in a journal Co-Citation network (CC), where the nodes of the network are journals and the (weighted) edges are co-citations representing the number of times two journals are jointly cited \cite{RN26}. Co-citations can be considered as a window on the intellectual world of journals: two journals having many co-citations can be considered as belonging to a same intellectual environment, such as a research field or sub-field \cite{Khelfaoui}. The second layer contains information about authors publishing articles in journals. The information is organized as an Interlocking Authorship network (IA), where nodes are journals and the (weighted) edges linking a pair of journals indicate the number of authors publishing in both of them \cite{RN35, RN36, RN40}. Authors of a journal represent the scholarly community gathered around this journal. Indeed, the basic idea consists in using information about communities of authors for individuating the intellectual proximity of journals, since communities of authors reflect to a certain degree the content of the journals and the choices of their editorial boards \cite{Brogaard,RN40}. The third layer deals with editors of journals. The information is organized as an Interlocking Editorship network (IE), where nodes are journals and the (weighted) edges linking a pair of journals indicate the number of their common editors  \cite{RN21}. Similarly to the IA layer, the focus is on scholarly communities gathered around the journals. In this case, communities represent groups of gatekeepers of journals. Actually, if two journals share persons on their editorial boards, they are likely to have similar editorial policies or similar paper selection processes, since they are managed by similar groups of people. 

These three layers of information were separately explored in a previous paper \cite{RN40} by means of a two-step analysis. In the first step, journals were classified by using a clustering algorithm applied to each layer. In the second step the coherence among the classifications obtained in the three layers were assessed by adopting a generalized correlation index.
The new techniques used in this paper consist in synthesizing the information about journals available in the three layer by building a new network, where information coming from each layer is combined and fused into a single layer. The fusion process is completely unsupervised and leverages on the structural properties of each layer. Subsequently, distance correlation and partial distance correlation indexes are used as indicators for estimating to what extent the information contained in each layer contributes to the determination of the fused network topology. Finally, a clustering algorithm is applied to the fused network in order to obtain a classification of journals based on the whole set of information available.

The paper is organized as follows. In Section 2 a short review of relevant literature is provided. Section 3 describes the network data. Section 4 is devoted to a short summary of the similarity network fusion technique and to the analysis of the contribution of each network to the fused network.
In Section 5 modularity is adopted for partitioning the fused network in communities of journals.
Section 6 contains the discussion of the results and concludes by suggesting further steps of the present research.

\section{A short review of literature}

This paper is at the crossroads between the literature devoted to the bibliometric delineation of scientific fields and the one dedicated to journal classification systems. Indeed, the characterization of scientific fields and of scholarly communities has a very long tradition. According to Borgman \cite[p. 589]{Borgman1989}, in bibliometric studies of scholarly communication, ``the most commonly asked research questions are of the form, What is the scholarly community of X? and, Of what types of scholars is the community composed?''. 
Zitt and coauthors \cite{Zitt2019}, in a recent review, highlighted that the question of the bibliometric delineation of scientific fields is inextricably interwoven with the definition of invisible colleges. Indeed, they consider the literature on delineation of scientific fields at the crossroads of two concepts: ``disciplinarity, which crystallizes various scientific activity in epistemology and sociology'' and ``invisible colleges, [...] the study of networks created explicitly or implicitly by publishing actors'' \cite[, p. 26]{Zitt2019}.  

From a theoretical point of view, the delineation of scientific fields or of scientific communities, and also the classification of journals, consists in partitioning the object of the analysis -- scholars, articles, journals --  in groups by using some classification technique. Zitt and coauthors \cite{Zitt2019} distinguished three different sets of techniques: (i) ready-made or institutional classifications of science which originate from scientists or librarians and do not entail the use of bibliometrics; (ii) bibliometric and scientometric networks resulting in groups of actors or artifacts. In this case,  discipline or subdiscipline labels are defined \textit{ex-post} by expert supervisors; (iii) information retrieval search, where classification relies on highly supervised schemes that are functional to efficient information retrieval.
 
By and large, the literature devoted to scientific fields delineation or to journal classification through the analysis of bibliometric and scientometric networks mainly uses single layer networks. This means that only one type of information goes into delineating a classification or into the definition of a science map \cite{Petrovich2020}. However, at least from Borgman \cite{Borgman1989}, scientific fields and scholarly communities are  defined by considering three ``theoretical variables'': producers of communication, artifacts of communication and concepts. A single layer approach uses these theoretical variables only one at a time.
 A big part of the literature focused on delineating scientific fields operationally defines ``articles'' as the ``artifacts''. Artifacts can be  linked by similarity of contents (word profiles; keywords), of citation profile (direct citations; co-citations; bibliographic coupling etc.), or of social attributes (authors). 
 There are many works aimed at comparing scientific field classifications emerging from the use of different proxies of similarities \cite{Boyack_Klavans_2010, Klavans_2017, Kleminski_2020}, or at evaluating the accuracy of the level, documents or journals, at which they are defined \cite{Shu_2018,Singh_2020}. 
 Other works developed journal classification systems based on single layer networks and compared them with ``ready-made classifications'' such the ones of  Web of Science or Scopus. For instance, Leydesdorff and coauthors \cite{Leydesdorff_2016} developed a classification of journals based on journal-journal citations. Wang and Waltman \cite{WangWaltman} developed a classification of journals based on direct citations. Khelfaoui and Gingras \cite{Khelfaoui}  applied the Louvain algorithm to a single-layer co-citation network of physics journals and found that the network was coherently structured by distinct, tough interconnected, communities of journals. Katchanov et al. \cite{Katchanov} also proposed a classification of physics journals by working on a single-layer similarity network based on citations.

Scientific fields and scholarly communities delineation is often reached starting from data about ``producers'' operationally defined as ``authors''. In a single layer perspective, node of the networks are the authors who are linked by social or intellectual relations, such as having collaborated as  co-authors \cite{Newman, Kretschmer}, or as having published articles in a same journal \cite{RN36, Brogaard, Carusi}. The application of tools for individuating clusters of authors, such as the community detection algorithms, partitions authors in groups interpreted as invisible colleges \cite{Price, Crane, Zuccala}. Clusters of authors are used also for delineating research fields and classifying journals \cite{RN36,Carusi}.

Literature focused on journal editors considers ``editors'' as a sort of ``producers'', more properly as \textit{gatekeepers of science} \cite{Crane, Braun}. The relations among editors consists in sitting together in the board of a journal. When tools for individuating clusters are applied to the network of editors, the resulting groups are interpreted as a kind of invisible college \cite{RN5, Goyanes}. This structure can be explored also in an interlocking editorship perspective: in this kind of network nodes are journals and the similarity between a pair of journals is computed by using the number of their common editors. The application of clustering tools to this network individuates groups of journals, the classification of which may be interpreted in terms of scientific fields delineation \cite{RN2, RN21, RN4,Lu,Teixeira,Xie_2020}. 

Until now, only single-layer approches to scientific fields delineation and classification of journals have been considered. But there are also contributions that use a multi-layer approach, i.e. that consider multiple information structured in more than one network.  A recent review of these works is provided by Zitt and coauthors \cite{Zitt2019} who labelled them as ``hybridization'' or ``multinetwork approches'', i.e.  contributions combining textual, citation and publishing actor networks. Their review is especially focused on the combination of textual and citation information. They clearly highlight that the use of multiple networks requires the heavy intervention of researchers for comparing or integrating different information contained in different networks. By considering only one example, the update of the UCSD (University of California San Diego) map of science \cite{Borner} combines data from Web of Science and Scopus. The final result was a map of the scientific fields and a corresponding classification of scholarly journals. It was built by integrating citation data and keywords data. The data gave rise to many journal-journal similarity matrices that were combined by using a completely ad-hoc supervised procedure, in which researchers defined not only the parameters for the specific algorithms used, but also the weights adopted for extracting or combining information.

The problem of combining or synthesizing information is even more complex when citation networks are considered together with actor networks. Indeed, there is the theoretical problem of merging cognitive structure and information about invisible colleges. 
Two articles made an attempt to combine multiple information for a fine grained delineation of scientific fields and a consequent classification of journals. 
Ni, Sugimoto and Cronin \cite{RN35} considered four networks of journals generated by common authors (producers), co-citation (artifact), common topics (concept) and common editors (gatekeepers). They searched for clusters in each of the networks and they observed if there are sets of journals grouping together across all the networks. Finally, they used a quadratic assignement procedure for comparing the proximities between journals in the four networks, and found that editors, authors and co-citations, but not topics, generated similar journal proximity networks.
The contribution of Baccini and co-authors \cite{RN40}, as already summarized in the introductive section, represents an improvement from a methodological point of view. It adopted a two-step analysis. In the first step, journals were classified by using a community detection algorithm to each of the three considered layers (authors, co-citations, editors). The application of the algorithm resulted in three different classification of journals. The second step of the analysis consisted in a statistical analysis of the coherence among the classifications obtained in the three layers by adopting a generalized correlation index.

\section{Journal networks data}

The present paper is actually a deeper analysis of the dataset used in  \cite{RN40}.  The dataset deals with the journals indexed in the Web of Science for three research fields, i.e. economics, information and library sciences and statistics in a selected period (see Table 1). As preliminarily stated in Section 1, connections between journals are represented by co-citations, common authors and common editors. Hence, the dataset may be efficiently rephrased as a multilayer network. For a comprehensive introduction to this specific topic, see \cite{Bianconi, dickison_2016}, while for a general introduction to networks see \cite{Newman}. Multilayer networks are used when a set of $n$ nodes is connected by links indicating $m$ different types of interactions -- which represent the $m$ layers. In a graphical setting, if a different colour is adopted for each type of connection, a multilayer network may be displayed as a unique coloured network. As an example, a social network can be represented by a multilayer in which the same group of people (the nodes) is connected via email (the first layer) or via mobile phone (the second layer). 

In the present dataset, a multilayer network with $m=3$ layers is considered for each scientific field. In such a case, nodes of each multilayer network represent journals. In the first layer, the classical CC layer, two journals are tied when they are both cited in the same paper. In the IA layer, two journals are linked when a scholar is the author of a paper in both journals. In the IE layer, two journals are linked if they have at least one scholar sitting in both their editorial boards. This three-layer network is properly a multiplex, since all the layers share exactly the same nodes.

A complete description of the dataset can be found in \cite{RN40}. Moreover, the raw data for the three multilayer networks for each discipline can be downloaded from the site \url{https://doi.org/10.5281/zenodo.3350797}. Table 1 reports the numbers of journals considered for each field and the surveyed years.

\begin{table}{h}
\caption{Research fields and surveyed periods for the considered networks.}
\centering
\scriptsize
{}
{%
\begin{tabular}[t]{ccccc}
\hline
Field&Number of journals&CC&IA&IE \\
\hline
Economics &169&2006-2010&2006-2010&2006\\
&&&&\\
Information and library sciences&59&2010-2014&2010-2014&2010\\
&&&&\\
Statistics&79&2006-2010&2006-2010&2006\\
\hline
\
\end{tabular}%
}
\begin{tablenotes}
\item []Source: \cite{RN40}. Raw data available from https://doi.org/10.5281/zenodo.3350797. 
\end{tablenotes}{}
\end{table}

\section{Similarity network fusion}

In order to synthesize the information contained in the multilayer network for each discipline, i.e. in order to collapse the multilayer network into a single network, we adopt the technique based on the Similarity Network Fusion (SNF), as proposed by \cite{wang2,wang2014}. SNF is an unsupervised method that can be used in many different application fields when several measurements collected on a set of units need to be integrated and studied in terms of similarity (for an application in bio-informatics, see \cite{baccini2020graph}). SNF is actually based on a given number of similarity matrices, corresponding to the various layers, referring to the same set of nodes. The similarity matrices are then integrated into a unique similarity matrix by means of the Cross Diffusion Process (CDP) \cite{wang2,wang2014}. CDP is an unsupervised iterative procedure that has the aim of reinforcing very strong links that are present in the single layers, as well as weaker links that are common to all the layers. The nodes of the resulting network are the common nodes of the layers, while the edges are weighted according to the new similarity values obtained through CDP. 

Let us suppose that $m$ different layers are considered for a set of $n$ nodes. Let $S^{(l)}=(s_{ij}^{(l)})$ denote the similarity matrix of order $n$ for the $l$-th layer, where  $l=1,\ldots,m$. In addition, let us consider the matrix $P^{(l)}=(p_{ij}^{(l)})$ obtained by normalizing the similarity matrix  $S^{(l)}$, i.e.
\begin{equation}\
p_{ij}^{(l)}=\frac{s_{ij}^{(l)}}{\sum_{g,h=1}^{n}s_{gh}^{(l)}}\ . 
\end{equation}
Moreover, we define a “local” similarity matrix $Q^{(l)}=(q_{ij}^{(l)})$, where
\begin{equation}\label{locsim}
q_{ij}^{(l)}=\begin{cases} \frac{s_{ij}^{(l)}}{\sum_{h\in N_i}s_{ih}^{(l)}}, & j\in N_i
\\
0 &$otherwise$
\end{cases}
\end{equation} 

In expression (\ref{locsim}), $N_i$ denotes the set of the $k$-th nearest neighbours of the $i$-th node -- computed by using the $k$-Nearest Neighbors ($k$-NN) algorithm (for more details on this technique, see e.g. Hastie et al., 2009). It should be remarked that the parameter $k$ is fixed and has to be specified in advance. The $m$ local similarity matrices essentially contain measures of the similarity between a node and its most similar neighbours. Indeed, the similarity between non-neighbouring nodes is set to $0$ in each matrix $Q^{(l)}$. 

The core of CDP is an iterative procedure which updates the “status” matrices at each step. The matrices $P^{(1)},\ldots,P^{(m)}$ are named “initial status” matrices and they constitute the input for CDP. The updating of a status matrix consists in combining the information contained in the local similarity matrix $Q^{(l)}$ with the information given by the normalized similarity matrices $P^{(h)}$ for $h\neq l=1,\ldots,m$. Indeed, if $P_{t}^{(l)}$ denotes the $l$-th status matrix computed at the $t$-th iteration with $P_{0}^{(l)}=P^{(l)}$, the updating rule for the status matrices is defined as
\begin{equation}\label{update}
P_{t+1}^{(l)}=Q^{(l)} \left(\frac{1}{m-1}\sum_{h\ne l=1}^m P_{t}^{(h)}\right)  (Q^{(l)})^\text{T}\ ,
\end{equation}
for $l=1,\ldots,m$.
The updating rule reflects the idea that local similarities of a single layer are combined with an average status matrix of the remaining layers. Finally, 
the corresponding overall status matrix is computed as
\begin{equation}\label{update}
P_{T}=\frac{1}{m}\sum_{l=1}^m P_{T}^{(l)}\ ,
\end{equation}
where $T$ represents a suitable iteration number when convergence of the algorithm is achieved (for more details on the choice of $T$, see \cite{wang2}). 

\begin{sidewaysfigure}
 \centering	
     \includegraphics[width=\textwidth]{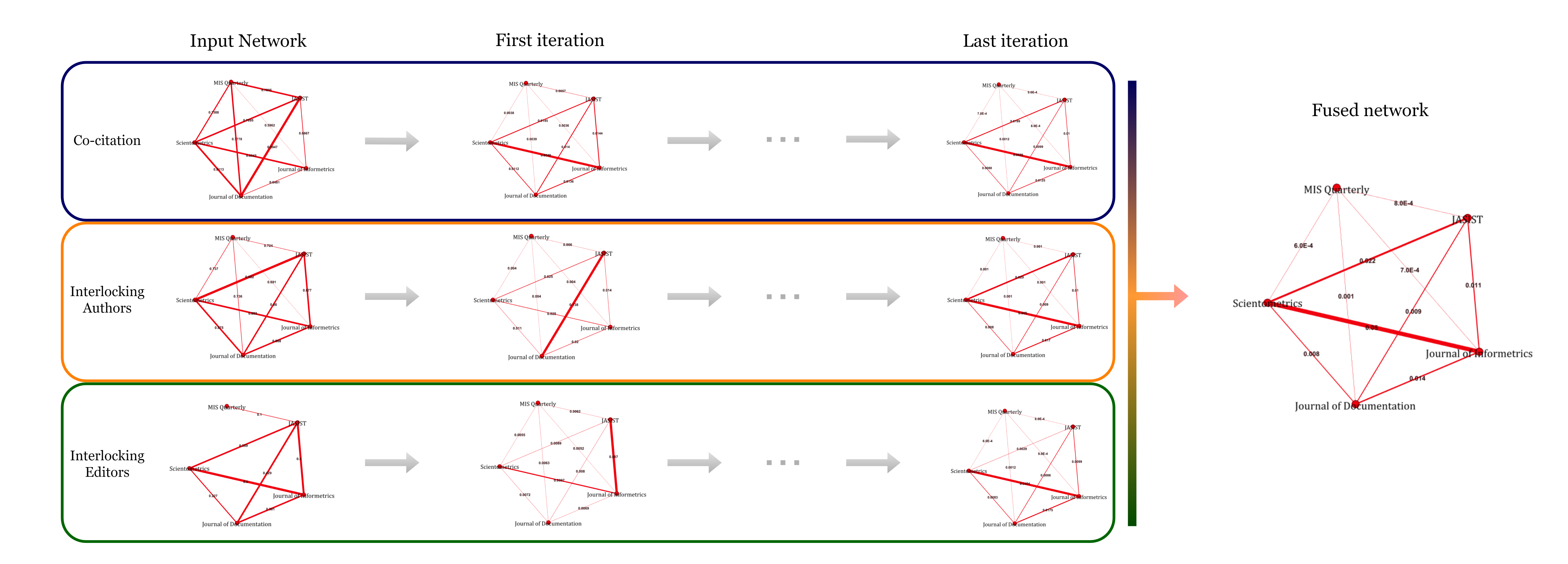} 
   \caption{Some phases of the application of SNF to subgraphs of the CC, IA and IE networks of five journals from the area of Information and Library Sciences. In the graphs, vertices correspond to journals, and edges are labeled with the value of the similarity between neighboring nodes. Starting from the left, the input subnetworks are depicted. Moving to the right, the graphs after the first iteration of CDP are reproduced. They are followed by the networks after the last iteration of CDP. Finally, the output network is depicted.}
   \label{fig:snfout}
\end{sidewaysfigure}

Intuitively, the iterative procedure has the aim of enriching the information of the single layers with that coming from the other layers. Indeed, densely connected groups of nodes are mantained and enforced with tighter links by SNF. At the same time, low weighted links that are present in all the layers become more visible with SNF, as they potentially represent stable relationships among groups of objects. In order to highlight this intuition, Figure \ref{fig:snfout} schematizes the application of CDP to a small subnetwork of five journals in the area of Information and Library Sciences consisting of \textit{JASIST}, \textit{Journal of Documentation}, \textit{Journal of Informetrics}, \textit{MIS Quarterly} and \textit{Scientometrics}. In the graphs, nodes correspond to journals, and edges are proportional to the similarity between linked nodes. In particular, starting from the left, the three input graphs, corresponding to the CC, IA and IE layers, are depicted. Moving one step to the right, the networks resulting from the first iteration of CDP are reported. Following the direction of the arrows, we then report the networks configuration at the last iteration of CDP, followed by the output network. It can be observed, for instance, that the link between \textit{Scientometrics} and \textit{Journal of Informetrics} is progressively reinforced by the fusion process. This is due to the presence of a strong link between the two journals which characterizes all the CC, IA and IE networks. We can also observe that the link between \textit{MIS Quarterly} and \textit{Scientometrics} is absent in the IE network. However, it has a non-zero weight in the output, as the two journals are tightly linked in the CC layer. Finally, the clique formed by \textit{Scientometrics}, \textit{Journal of Informetrics} and \textit{JASIST}, becomes more evident in the output network, while it is present, but less visible, in the single layers. The above observations on Figure \ref{fig:snfout} reflects the property of SNF of reinforcing strong and common links.

The software for the practical implementation of SNF may be downloaded in the \proglang{R} or \proglang{MATLAB} version at the site \url{http://compbio.cs.toronto.edu/SNF/SNF/Software.html}

In the present application of SNF, $m=3$ layers were considered for each research field, i.e. economics, information and library sciences, and statistics. Hence, three similarity matrices were computed for each layer corresponding to CC, IA and IE. Each similarity matrix was implemented on the basis of
the classical Jaccard similarity coefficient between sets \cite{Jaccard}. We remind that if $A$ and $B$ represent two sets, the Jaccard coefficient is defined as
\begin{equation}
J(A,B)=\frac{| A \cap B |}{| A \cup B |}\ ,
\end{equation}
where $\mid \cdot \mid$ denotes the cardinality of a set. Obviously, it holds $0\le J(A,B) \le 1$. As an example, if the IA layer is considered, the similarity between two journals is proportional to the number of authors writing papers in both the journals. Hence, if two journals have exactly the same set of authors, i.e. when $A=B$, the maximum similarity $J(A,B)=1$ is achieved. In contrast, the minimum similarity $J(A,B)=0$ occurs when two journals have no authors in common, i.e. when $A \cap B =\emptyset$.
Thus, the entries of the matrix $S^{(l)}$ were suitably computed as Jaccard coefficients for each layer. The layers were subsequently fused into a unique network through CDP by setting $k=5$ and $T=10$. Similarity matrices for each layer and for the fused networks of the three reasearch fields are available from \url{https://doi.org/10.5281/zenodo.4537205}. Different choices of these parameters gave rise to similar fused networks, i.e. the algorithm is rather stable. The implementation of SNF was done in the \proglang{R} environment by using the \code{SNF} function in the \pkg{SNFtool} package (R Core Team, 2020). The obtained fused networks for each discipline are displayed in Figures 2-4.

In order to assess the relation of the network obtained by means of SNF with the original layers, the generalized distance correlation $R_d$ suggested by \cite{szekely2007} was considered. The index may be adopted in order to compare the dependence between similarity matrices, as also suggested by \cite{RN23} in an environmental setting. The interpretation of the generalized distance correlation is similar to the squared Pearson correlation coefficient. Hence, the quantity $\sqrt{R_d}$ may be seen as a generalization of the usual correlation coefficient. Actually, $R_d$ is defined in the interval $[0,1]$, in such a way that values close to zero indicate no or very weak association, while larger values suggest a stronger association, which is perfect for $R_d=1$ - similar considerations obviously hold for $\sqrt{R_d}$ (for more details, see \cite{RN23}). 

In the present framework, the partial distance correlation $R_d^*$ proposed by Székely and Rizzo \cite{székely2014} is also considered. This further index permits to measure the degree of association between two similarity matrices corresponding to the fused network and a layer, with the effect of a further layer removed. Indeed, if the focus is on the relationship between the fused network and a layer, using their distance correlation may provide misleading conclusions if there is a further confounding layer which is related to the fused network. This misleading information can be avoided by controlling for the confounding layer and -- hence -- by adopting the partial distance correlation. The generalized and partial distance correlations were evaluated in the \proglang{R}-computing environment (R Core Team, 2020) by using the \proglang{R} functions \code{dcor} and \code{pdcor} in the package \pkg{energy}. The computed values of these coefficients are reported in Tables 2-4 for the three scientific areas.

\begin{table}[!h]
\centering
\scriptsize
\caption{Generalized and partial distance correlation for the field of economics. F denotes the fused network. The first row reports the value of generalized distance correlation between F and the layers. The subsequent rows report the values of partial distance correlation between F and the layers, conditioned on each single layer.}
\bigskip
\begin{tabular}{ccccc}
\hline
economics & & CC & IA & IE\\
\hline
F & $\sqrt{R_d}$ & 0.4467 & 0.4483 & 0.8079\\ 
& & & & \\
F$|$CC & $\sqrt{R_d^*}$ & - & 0.1047 & 0.5975 \\ 
& & & & \\
F$|$IA & $\sqrt{R_d^*}$ & 0.1901 & - & 0.6031  \\ 
& & & & \\
F$|$IE & $\sqrt{R_d^*}$ & 0.1944 & 0.1516 & -  \\ 
& & & & \\

\hline		
\end{tabular}
\label{econ}
\end{table}

\begin{table}[!h]
\scriptsize
\centering
\caption{Generalized and partial distance correlation for the field of information and library sciences. F denotes the fused network. The first row reports the value of generalized distance correlation between F and the layers. The subsequent rows report the values of partial distance correlation between F and the layers, conditioned on each single layer.}
\bigskip
\begin{tabular}{ccccc}
\hline
information and library sciences & & CC & IA & IE\\
\hline
F & $\sqrt{R_d}$ & 0.6167 & 0.5997 & 0.7767 \\ 
& & & & \\
F$|$CC & $\sqrt{R_d^*}$ & - & 0.1766 & 0.7013 \\ 
& & & & \\
F$|$IA & $\sqrt{R_d^*}$ & 0.2601 & - & 0.7053 \\ 
& & & & \\
F$|$IE & $\sqrt{R_d^*}$ & 0.2985 & 0.2529 & - \\ 
\hline		
\end{tabular}
\label{econ}
\end{table}

\begin{table}[!h]
\centering
\scriptsize
\caption{Generalized and partial distance correlation for the field of statistics. F denotes the fused network. The first row reports the value of generalized distance correlation between F and the layers. The subsequent rows report the values of partial distance correlation between F and the layers, conditioned on each single layer.}
\bigskip
\begin{tabular}{ccccc}
\hline
statistics & & CC & IA & IE \\
\hline
F & $\sqrt{R_d}$ & 0.5349 & 0.5202 & 0.7898 \\
& & & & \\
F$|$CC & $\sqrt{R_d^*}$ & - & 0.1760 & 0.5275 \\ 
& & & & \\
F$|$IA & $\sqrt{R_d^*}$ & 0.2348 & - & 0.5305 \\ 
& & & & \\
F$|$IE & $\sqrt{R_d^*}$ & 0.1947 & 0.1372 & - \\ 
\hline		
\end{tabular}
\label{econ}
\end{table}

Tables 2-4 show that the fused network is generally associated with the three layers for each discipline, as indicated by the values of the generalized distance correlations between the similarity matrix of the fused network and the similarity matrices of each layer.
Moreover, the generalized distance correlation is the highest for the three research fields when the IE layer is considered, with values near to 0.8. This may suggest that the IE contribute the most in structuring the fused networks in the three research fields. 
As anticipated, the high value of the generalized distance correlations between the fused network and the IE layer may be confounded by the association between the fused network and the other two layers. The estimates of the partial distance correlation permits to evaluate the degree of association between the fused network and the IE layer, by removing the confounding effect of the other layers. 
Consider for example Table 2. In economics, the generalized distance correlation between the fused network and the IE layer is 0.8079.
This value is confounded by the association between the fused network and the IA and CC layers. The partial distance correlation F$|$CC (F$|$IA) indicates the association between the fused network and the IE network after removing the confounding effect represented by the association between the fused network and the CC (IA) layer. The estimated partial distance correlation assume lower values when the confounding effect of the CC and IA layer is removed (0.5975 and 0.6031, respectively).

Analogously, the values of the generalized distance correlation between the fused network and the CC and IA layers are similar and near to 0.5. The partial distance correlation assumes much lower values, indicating that the degree of association between the fused network and the CC and IA layer is much lower when the confounding effect of the other layers is removed.

In sum, Tables 2-4 indicate that the estimated values of the partial distance correlations are high solely for the IE layer in the three research fields, i.e. the fused network is specially associated with the IE layer. If the degree of association is interpreted as an indicator of the contribution of each layer to the structure of the fused network, it is possible to say that the IE layer contribute the most to the structure of the fused network in the three research fields. The information and library science field is charcaterized by the highest value of the partial distance correlations for the IE layer, followed by the economics field.

\section{Modularity and clusters of journals}

The similarity network fusion process permits to synthesize different layers of information in a unique network. The use of generalized and partial distance correlation allows to determine the contribution of each layer of information to the structure of the fused network. Hence, it is possible to search and to visualize communities in a unique network by adopting usual network analysis techniques. 

In the present setting, each fused network synthesizes information about intellectual and social similarity among journals. The application of network fusion techniques to explore multi--level similarity among journals represents a major improvement with respect to previous works \cite{RN35,RN40}, where clustering or community detection algorithms were applied separately to different layers. As a matter of fact, multiple classifications of journals emerged for a same set of journals, which have to be evaluated by qualitative inspection, as in \cite{RN35}, or by using statistics, as in \cite{RN40}. Moreover, multiple classifications make it difficult to interpret the results.

In this work, for each research field, the fused network is partitioned by using both the Louvain algorithm \cite{RN37} and the VOS (Visualization Of Similarities) algorithm as implemented in the software  \proglang{Pajek} \cite{RN15}. The Louvain algorithm consists in the optimization of the modularity of the network \cite{RN39,RN38}, while the VOS algorithm optimizes a measure called VOS quality. Table 5 shows the resolution parameters adopted for optimization, the number of detected communities and the values of modularity and VOS qualities. For each scientific field, results of the two algorithms are similar, and the corresponding classification of journals are identical. 

\begin{sidewaysfigure}
 \centering	
     \includegraphics [scale=0.3]{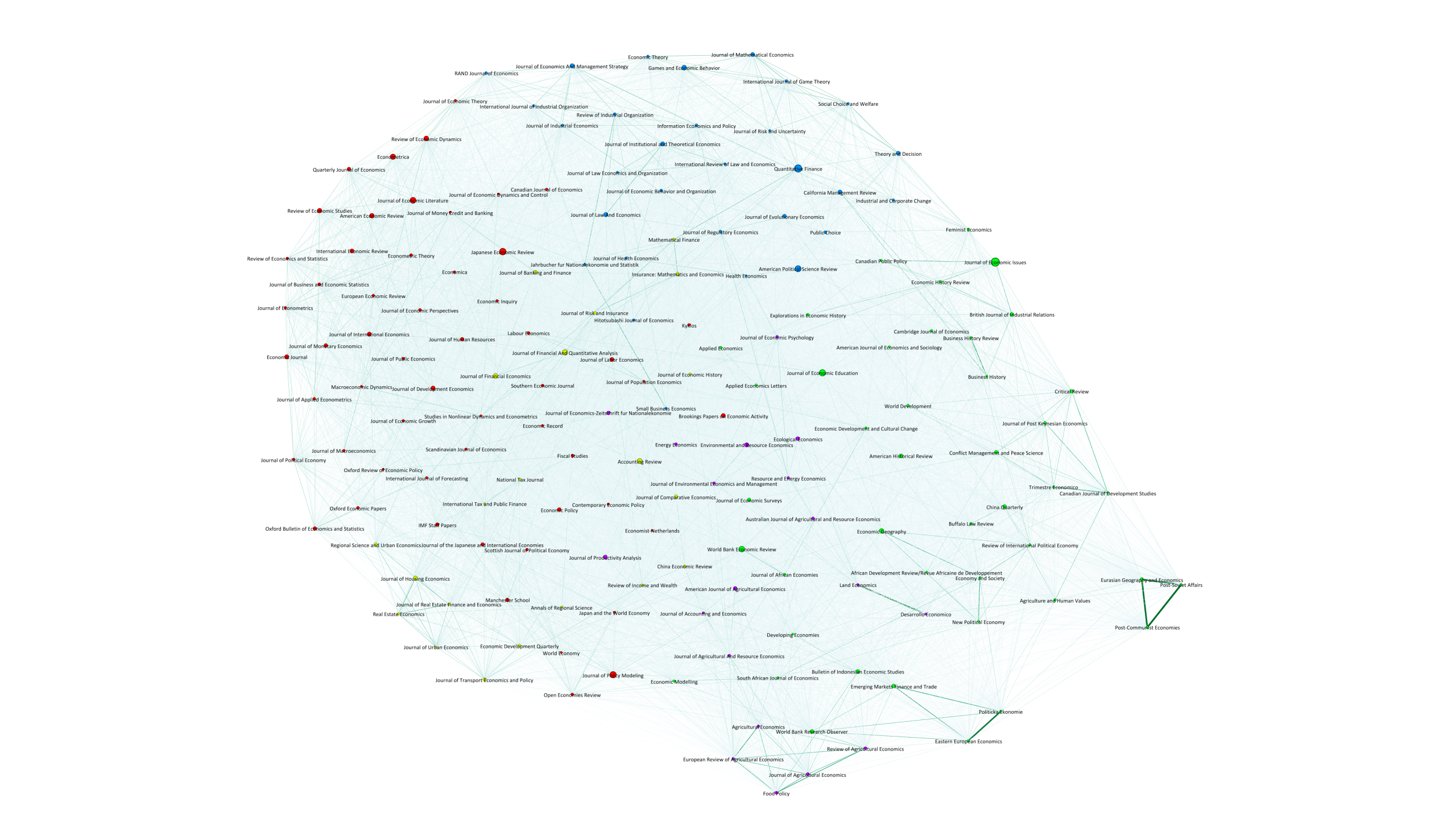} 
   \caption{Economics: the network of journals. Different colours indicate different clusters. Size of a node is proportional to its impact factor.Thickness of an edge is proportional to the similarity score between its endpoints in the fused network.}
   \label{fig:IE_ECON}
\end{sidewaysfigure}

\begin{sidewaysfigure}
 \centering	
     \includegraphics [scale=0.3]{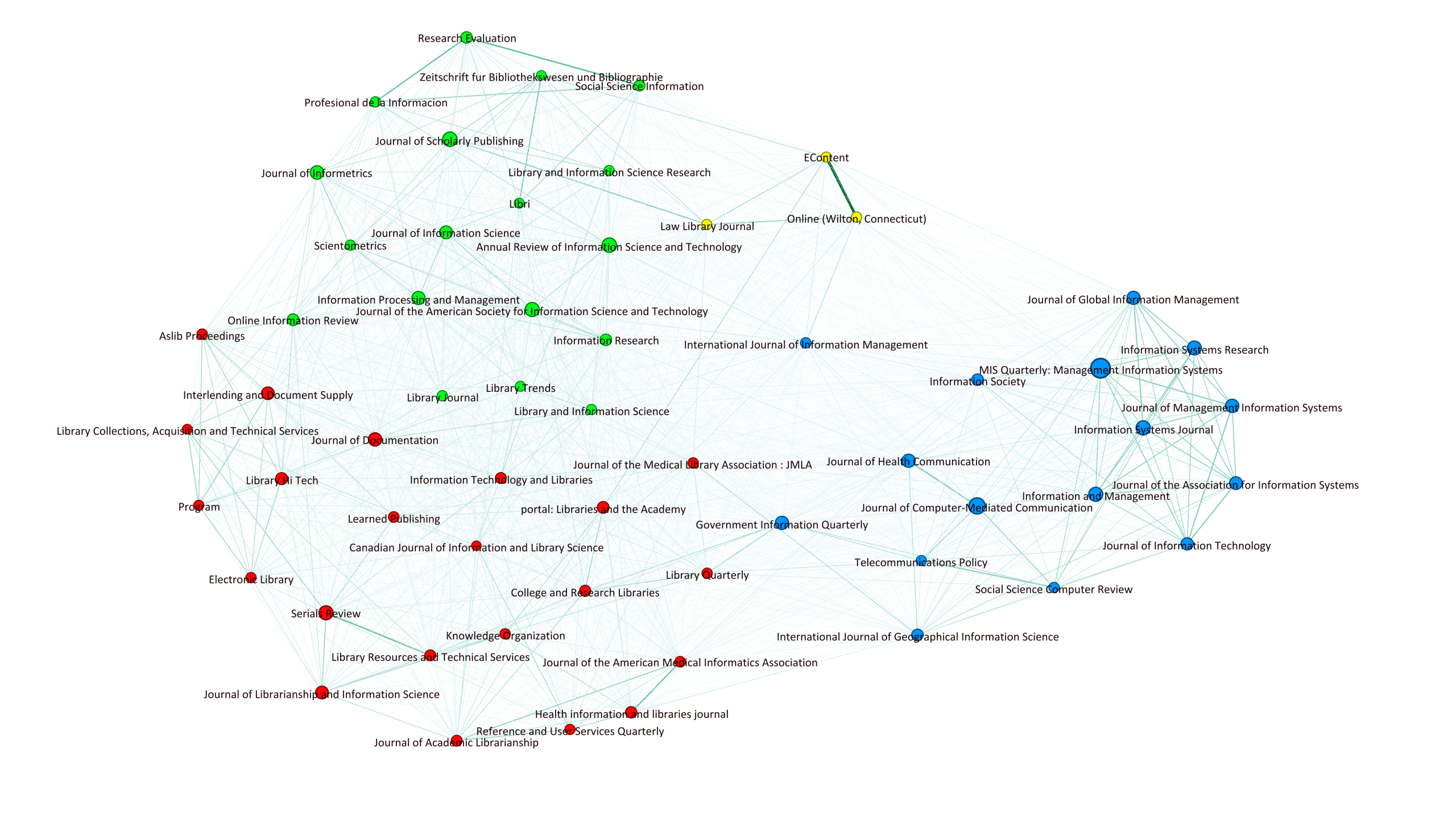} 
   \caption{Information and library science: the network of journals. Different colours indicate different clusters. Size of a node is proportional to its impact factor.Thickness of an edge is proportional to the similarity score between its endpoints in the fused network.}
   \label{fig:IE_ILS}
\end{sidewaysfigure}

\begin{sidewaysfigure}
 \centering	
     \includegraphics [scale=0.3]{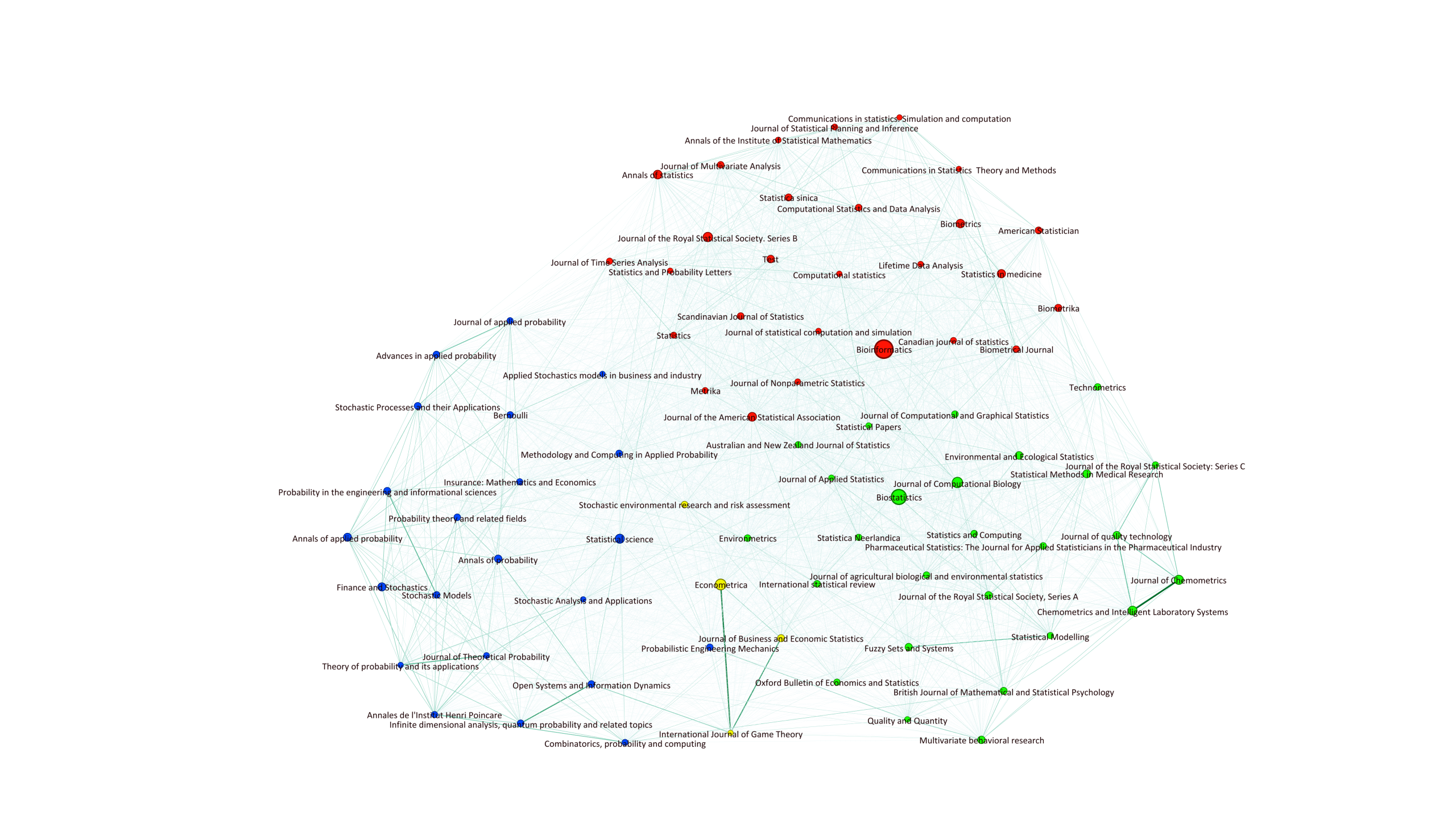} 
   \caption{Statistics: the network of journals. Different colours indicate different clusters. Size of a node is proportional to its impact factor.Size of a node is proportional to its impact factor.Thickness of an edge is proportional to the similarity score between its endpoints in the fused network.}
   \label{fig:IE_STAT}
\end{sidewaysfigure}

The fused networks for the three fields are displayed in Figures \ref{fig:IE_ECON}-\ref{fig:IE_STAT}, where different node colors indicate different clusters of journals. Size of nodes is proportional to the Impact Factor (IF) of journals, as registered in the \textit{Journal Citation Reports} for the year corresponding to the one in which the IE was surveyed. The lists of journals, their classification and IF are available at \url{https://doi.org/10.5281/zenodo.4537205}. Thickness of an edge is proportional to the similarity score between its endpoints in the fused network. Network data are energized by using the ForceAtlas2 algorithm \cite{forceatlas} of the software \proglang{Gephi} \cite{gephy}. The algorithm turns structural proximities into visual proximities, and communities appear as groups of near nodes.
 
For the field of economics, five communities are detected. The community with red nodes groups 53 journals that can be considered the core of the mainstream way to do economics. This group appears to be characterized by a strong presence of journals devoted to macroeconomics, international economics and monetary economics. It also contains the red general purpose journals, with the highest IF, such as the \textit{Journal of Economic Literature} or the \textit{Economic Journal}, and, among the others, all the so-called top five-journals of economics \cite{Heckman}: \textit{The American Economic Review}, \textit{Econometrica}, the \textit{Journal of Political Economy}, the \textit{Quarterly Journal of Economics} and the \textit{Review of Economic Studies}. In this groups there are also the top journals of some specialized fields, such as the \textit{Journal of Development Economics}, the \textit{Journal of Economic Growth} and the \textit{Journal Financial Economics}.

The blue node community gathers 33 journals mainly devoted to economic theory, game and decision theory. It covers also journals with specific applications of microeconomics, such as law and economics, health economics and industrial economics. It contains also a couple of journals mainly devoted to economics of innovation and listed as 'heterodox' by the \textit{Heterodox Economics Directory}: the \textit{Journal of Evolutionary Economics} and \textit{Industrial and Corporate Change} (http://heterodoxnews.com/hed/journals.html). In this case, the specialization appears to win over the methodological approach: a suitable label for this cluster of journals is ``mainstream microeconomics''.  

The yellow node community collects 20 journals specialized in various applied fields, such as public economics (e.g. \textit{Fiscal Studies, National Tax Journal}), regional urban economics, labor economics and financial economics. With a few exceptions, these journals have lower impact factors than the two previously commented communities. These journals can be considered properly as specialized journals.

The purple node community gathers 21 specialized journals on agricultural, environmental and energy economics. In this group there are also journals at the boundary between economics and other fields, such as accounting.

Finally, the green node community groups 42 journals that appear to be either clearly heterodox or characterized by methodological pluralism. The only general purpose journal of this group is the \textit{Cambridge Journal of Economics}, which is well-known for accepting also non-mainstream contributions. Other journals are devoted to heterodox approaches such as \textit{Journal of Post-Keynesian Economics} or \textit{Feminist Economics}; others are focused on evolutionary economics such as \textit{Journal of Economic Issues}. In this group there are also other specialized journals in fields that are near to other disciplines, such as economic history or geography. 

It is very interesting to note that journals with the highest IF, i.e. the most popular journals of each cluster, are positioned by the graphic algorithm at the periphery of each network. As a consequence, the journals that have the highest IF within a cluster are the most dissimilar from the journals with the highest IF of the other clusters. For example, the red top-five journals of economics are very far from green heterodox journals such as the \textit{Journal of Post-Keynesian Economics} or the  \textit{Cambridge Journal of Economics}. 

As to the information and library science journals, 4 communities are detected, challenging the traditional bi-partition of the field in management of information and the library and information science \cite{RN40,Sugimoto_2008}. Indeed, the blue node community groups 16 journals mainly devoted to management of information. The red node community gathers 22 journals mainly devoted to 'traditional' library science. The green node community gathers 18 journals focused on or drifted toward scientometrics and informetrics, such as \textit{Scientometrics}, the \textit{Journal of the American Society for Information Science and Technology}, the \textit{Journal of Informetrics}. The three journals belonging to the yellow node community are professional journals for librarians. 

Finally, statistical journals are grouped in 4 communities. The first community (blue nodes of the network in Figure \ref{fig:IE_STAT}) gathers 22 journals which are mainly devoted to probability theory and its applications. Indeed, it contains the leading journals in this scientific area, such as \textit{Annals of Probability}, \textit{Annals of Applied Probability}, \textit{Probability Theory and Related Fields} and \textit{Stochastic Processes and their Applications}.
Some reviews, such as \textit{Bernoulli} and \textit{Statistical Science}, are also embedded in this group, since they usually publish many papers with particular attention to themes of probability.

The second community (green nodes of the network in Figure \ref{fig:IE_STAT}) contains 27 journals primarily dealing with the issues of theoretical statistics. Typical members of this group are the “big four” statistical journals, i.e. \textit{Annals of Statistics}, \textit{Biometrika}, \textit{Journal of the American Statistical Association} and \textit{Journal of the Royal Statistical Society - Series B}. These reviews publish papers with strong methodological contents - often aimed at handling original and complex datasets. Interestingly, this community also contains journals (such as \textit{Biometrics} and \textit{Bioinformatics}) that - despite their applied nature - produce many technical articles. 

The third community (red nodes of the network in Figure \ref{fig:IE_STAT}) includes journals dealing with various topics of applied statistics. As a matter of fact, reviews in this group promote statistical applications to biology, ecology, chemistry and social sciences. Typical members of this community are \textit{Biostatistics}, \textit{Environmental and Ecological Statistics}, \textit{Environmetrics}, \textit{Chemometrics} and \textit{Multivariate Behavioral Research}.

Finally, the last community (yellow nodes of the network in Figure \ref{fig:IE_STAT}) constitutes a small component of the network, as it only contains 4 journals. However, this group has a very strong characterization, because it embraces the reviews at the boundary with the field of economics, such as \textit{Econometrica} and \textit{Journal of Business and Economic Statistics}. In conclusion, a precise and reasonable characterization can be attached to the communities obtained in the networks of the statistics area.

\begin{table}[!t]
\caption{Detected communities by using Louvain and VOS algorithms. For each research field the resolution adopted for tuning the community detection algorithm, the number of detected algorithms and the values of modularity and Vos quality are reported.}
\
\centering
\scriptsize

{}
\begin{tabular}{ccccccc}
\hline 
 & \multicolumn{3}{c}{Louvain algorithm} & \multicolumn{3}{c}{VOS algorithm}\\ 
\
\ &Resolution&Modularity&Communities&Resolution&VOS quality&Communities\\
\hline 
Economics&1&0.240&5&1&0.245&5 \\ 
\
Information and Library sciences&1&0.276&4&1&0.292&4 \\ 
\
Statistics&1&0.235&4&1&0.246&4 \\ 
\hline
\end{tabular} 

\end{table}


\newpage

\section{Discussion and conclusion}

The use of network structural information for classifying scientometric objects is not new. By contrast, the use of structural information about multi-layer networks to the same aim is relatively new. The basic idea of this paper is to build a unique journal network merging information about their cognitive structure and about invisible colleges gathered around them. This unique network was built by using a similarity network fusion technique that allowed to synthesize into a single network the information contained in a multi-layer network. Before the fusion, information was structured in a three-layers network. The first layer contained information about intellectual similarity among journals as proxied by co-citations. In the second layer similarity among journal was computed by considering the authors of the articles published in journals. The third layer, finally, considered the similarity between journals in terms of their editorial boards. The unsupervised similarity network fusion algorithm collapses the three layers of information in a single network where nodes are journals, and an edge is weighted according to the fused measure of similarity between its endpoints. 

In the fused network, similarity between journals was computed through an unsupervised algorithm that reinforces the strongest links present in a single layer, and the links that are common to all the layers. 

The analysis of the general and partial distance correlations permitted to show the relative contribution of each layer of information to the values of similarities in the fused network. According to our empirical data, the fused network appears to be especially associated with the IE layer. This result may be considered as suggesting that the role of editors acting as gatekeepers of science is particularly strong in determining similarity among journals. It appears that the similarity among journals reflects mainly the underlying social network in an academic field, rather than cognitive similarity among journals as measured by co-citations. As for economics, this result appears to be coherent with previous findings highlighting the role of editorial boards of journals \cite{RN21, RN5}. Heckman and Moktan suggested a possible explanation for that: ``low turnover in editorial boards creates the possibility of clientele effect surrounding both journals and editors, whereby authors, in an effort to increase their chance of publication, choose to conduct research that caters to the policy or methodology preferences of editors" \cite[, p. 422]{Heckman}. 

The last step of the analysis has consisted in searching for clusters of journals in the fused network by using standard modularity techniques. The possibility to search for communities in a single network where different information are fused together is a relevant improvement with respect to the alternative of detecting communities in different layers and studying the correlations among the detected clusters as in \cite{RN40}. This issue is true for all the fields considered in the present study, since the search for communities in the IE layer resulted in a higher number of journal clusters with respect to CC and IA layers \cite{RN40}. The existence of a different number of clusters in various layers introduces many difficulties, since it requests to interpret non-identical communities of journals in different layers as in \cite{RN35, RN40}. To combine \textit{ex-post} different classifications is a difficult task involving subjective choices about the weights to attribute to different information, such as in \cite{RN35}. The adoption of the similarity network fusion technique allows to overcome this drawback, by permitting to synthesize information contained in the various layers through an unsupervised algorithm before starting the detection of journals communities.

The empirical results are three fine-grained classifications of scholarly journals, one for each of the three scientific fields considered. The starting point of the analysis was a list of journals belonging to a scientific field according to the Web of Science classification. For each field, the final output is a classification of journals individuating similar groups from an intellectual (co-citations) and social (authors and editors) point of view. 

For information and library sciences, as well as statistics, the detected journal communities may be easily interpreted as a result of different and recognized sub-field specializations. In particular, for statistics, the resulting classification nearly completely overlaps with the sub-field classification suggested by \cite{RN2}, as a simple expert-based classification. For information and library sciences, the detected communities can be interpreted as the result of the ongoing evolution of the traditional bi-partition in management of information and the library and information science sub-fields.
For economics, instead, the detected communities appear to be also selected by different methodological approaches. Four groups of journals may be considered as representing different specialization inside the mainstream way to do economics, while a fifth community gathers outright heterodox journals, or at least pluralist journals that accept to publish also non-mainstream articles. The proposed methodology appears particularly promising, since it is able to detect clusters of journals originating from different academic dynamics, in our cases sub-field specialization for statistics and information and library sciences, and from different general approaches to research for economics. 

There are at least two natural extensions of the present analysis. The first is to add layers of information to the ones used here, such as topic or keywords. The second is about the possibility to generalize the results to other fields. In this paper, two fields belonging to social sciences -- economics and information and library sciences -- and one near to mathematics were considered. Is the prominent contribution played by IE network in defining the structure of the fused networks peculiar of the three fields considered here, or is it a general feature valid also for other fields? This is a key questions for deepening the understanding of the role of editors as gatekeepers of science. How much do the results found here depend on having considered fields with similar publishing strategies in terms of (limited) number of co-authors and similar editorial organization of journals? It would be very interesting to understand what is going on in fields with different habits such as life sciences, medicine and, especially, physics. In particular, how does a higher number of co-authors per paper impact on the structure of the IA layer? 
It would be very interesting also to verify the results in a diachronic perspective by considering different time periods for the same research field, as was done, for example, by Claveau and Gingras \cite{Claveau} for the field of economics between 1960s and 2000s using bibliographic coupling. 

The use of similarity network fusion may open interesting avenues for addressing any research question where information is structured as a multi-layer network.

\newpage

\printbibliography

\end{document}